\documentclass[11pt]{article}
\usepackage{amsmath, latexsym,ifthen, amsfonts,amsbsy}
\usepackage{setspace,subfig}
\usepackage{natbib}
\usepackage{relsize}
\usepackage{threeparttable}
\renewcommand{\harvardurl}[1]{\textbf{URL:} \url{#1}}
\usepackage{booktabs}
\usepackage{tkz-graph}
\usepackage{pgf}
\usepackage{setspace}
\usepackage{times}
\usepackage{lipsum}
\usepackage{subfig}
\usepackage{tabularx}
   
\usepackage{amsmath,amsthm,amssymb,ifthen}
\usepackage{graphicx}
\usepackage[mathscr]{eucal}

\usepackage{tikz}
\usetikzlibrary{arrows,shapes.arrows,shapes.geometric,
	shapes.multipartb,backgrounds,decorations.pathmorphing,positioning,fit,automata}
\tikzset{
	>=stealth',
		square/.style={regular polygon,regular polygon sides=4}
	true/.style={
		rectangle,
		draw=black, very thick,
		text width=6.5em,
		minimum height=2em,
		text centered,
		fill=gray, opacity = 0.5},
	punkt/.style={
		square,
		draw=black,very thick,
		text width=1.5em,
		minimum height=2em,
		text centered},
	est/.style={
		circle,minimum size=10mm,
		draw=black, very thick,
		text centered},
	shade/.style={
		circle,
		draw=black, very thick, fill=gray!50,
		text centered},
	weight/.style={
		circle,
		draw=black, very thick,
		text width=6.5em,
		minimum height=2em,
		text centered},
	pil/.style={
		->,
		thick,
		shorten <=2pt,
		shorten >=2pt,},
	double/.style={
		<->,
		thick,
		shorten <=2pt,
		shorten >=2pt,},
	dash/.style={
		dashed,
		thick,
		shorten <=2pt,
		shorten >=2pt,},
	dashdouble/.style={
		<->,
		dashed,
		thick,
		shorten <=2pt,
		shorten >=2pt,}
}
\usepackage{enumerate}

\usepackage{etex}
\usepackage{booktabs}
\usepackage{amsfonts}
\usepackage[latin1]{inputenc}
\usepackage{soul,color}
\usepackage{chngcntr}
\usepackage{amsthm}
\usepackage{verbatim}
\usepackage[hidelinks]{hyperref}
\usepackage{changepage}
\usepackage{amssymb}
\usepackage{adjustbox}
\newcommand{\bm}[1]{\mbox{\boldmath{$#1$}}}

\usepackage{caption} 
\usepackage{xfrac}
\usepackage{graphicx}

\newtheoremstyle{note}
{8pt}
{8pt}
{}
{}
{\bfseries}
{:}
{.5em}
{}

\theoremstyle{note}
\newtheorem{theorem}{Theorem}
\newtheorem{lemma}[theorem]{Lemma}

\usepackage[T1]{fontenc}

\allowdisplaybreaks

\usepackage{multirow}
\usepackage[left=1in,right=1in,top=1in,bottom=1in]{geometry}
\usepackage{algorithm}
\date{}
\usepackage{tikz}

\usepackage{soulpos}
\soulregister\cite7
\soulregister\citep7
\soulregister\ref7
\soulregister\emph7
\soulregister\eqref7
\soulregister\pageref7
\definecolor{mygreen}{RGB}{144,241,47}

\def\T{ {\mathrm{\scriptscriptstyle T}} }

\def\bSig\mathbf{\Sigma}

\usepackage[figuresright]{rotating}

\usepackage{authblk}
\author[]{Deshanee S. Wickramarachchi}
\author[]{Laura Huey Mien Lim}
\author[]{Baoluo Sun}
\affil[]{Department of Statistics and Data Science, National University of Singapore}

	{\raise0.5ex\hbox{$#1$}\! \left/ \! \lower0.5ex\hbox{$#2$}\right.}
	
\begin{document}

\title{Mediation Analysis with Multiple Mediators under Unmeasured Mediator-Outcome Confounding}

	\clearpage \maketitle
	

\begin{abstract}	
	{	
It is often of interest in the health and social sciences to investigate the joint mediation effects of multiple post-exposure mediating variables. Identification of such joint mediation effects generally require no unmeasured confounding of the outcome with respect to the whole set of mediators. As the number of mediators under consideration grows, this key assumption is likely to be violated as it is often infeasible to intervene on any of the mediators. In this paper, we develop a simple two-step method of moments estimation procedure to assess mediation with multiple mediators simultaneously in the presence of potential unmeasured mediator-outcome confounding. Our identification result leverages heterogeneity of the exposure effect on each mediator in the population, which is plausible under a variety of empirical settings. The proposed estimators are illustrated  through both simulations and an application to evaluate the mediating effects of post-traumatic stress disorder symptoms in the association between self-efficacy and fatigue among health care workers during the COVID-19 outbreak.
	
	}
\end{abstract}	

\section{Introduction}

Although evaluation of the total effect of a given exposure on an outcome of interest remains the goal of many empirical studies, there has been considerable interest in recent years to gain deeper understanding of  causal mechanisms by teasing apart the direct exposure effect and the indirect effect operating through a given post-exposure mediating variable \citep{vanderweele2015explanation}.  While mediation analysis  has a longstanding tradition primarily in the context of linear structural modeling \citep{ctx8536838640005154, bentler1980multivariate, baron1986moderator}, the concepts of natural direct and indirect effects have since been formalized under the potential outcomes framework by the seminal works of \cite{robins1992identifiability} and \cite{pearl2001direct}, which disentangles the identification assumptions from parametric constraints and allows semiparametric identification and inference for these causal estimands of interest \citep{imai2010identification, tchetgen2012semiparametric,tchetgen2014estimation, vanderweele2015explanation,hines2020robust}. 

Identifcation of natural direct and indirect effects generally requires no unmeasured confounding of the exposure-mediator, exposure-outcome and mediator-outcome relationships \citep{robins1992identifiability, pearl2001direct}. While the no unmeasured exposure-mediator and exposure-outcome confounding assumptions hold by design when the exposure is randomly assigned by the investigator, it is often infeasible to also intervene on the post-exposure mediator.  As a result, one can rarely rule out unmeasured mediator-outcome confounding in empirical settings and sensitivity analysis methods have been developed to assess the impact of departures from this assumption \citep{imai2010general,vanderweele2010bias, tchetgen2012semiparametric, ding2016sharp,zhang2022interpretable}.  A growing literature has likewise emerged on the development of  identification strategies in the presence of potential unmeasured mediator-outcome confounding, typically by leveraging an instrumental variable for the effect of the mediator on the outcome \citep{ten2007causal,doi:10.1002/sim.2891, albert2008mediation,small2012mediation, imai2013experimental,burgess2015network,zheng2015causal,doi:10.1111/rssb.12232,tai2021identification}. 

While most of the aforementioned literature has been restricted to the setting of a single mediator, researchers often focus on multiple mediators, either to assess the joint mediation effects through multiple mediators or address concerns about mediator-outcome confounding by exposure-induced intermediates \citep{vanderweele2014mediation,vanderweele2014effect}. Identification of such joint mediation effects require no unmeasured confounding of the outcome with respect to the whole set of mediators \citep{vanderweele2014mediation}. This key assumption is arguably more likely to be violated as the number of mediators under consideration grows, and the development of methodology to adequately address this issue remains a priority in mediation analysis. Nonetheless, extension of the instrumental variable identification approach to the multiple mediator setting presents some practical challenges as it generally requires as many instrumental variables as there are mediators. 

Recently, \cite{fulcher2019estimation} proposed an alternative identification approach by using higher order moment conditions which does not require ancillary variables such as instrumental variables \citep{rigobon2003identification, klein2010estimating, lewbel2012using, tchetgen2021genius}.  In this paper, we build upon and extend the works of  \cite{vanderweele2014mediation} and \cite{fulcher2019estimation} to the multiple mediator setting in the presence of potential unmeasured mediator-outcome confounding. The rest of the paper is organized as follows. We introduce notation and assumptions in Section \ref{sec:notation} and delineate the conditions for identification of mediation effects with multiple mediators simultaneously in Section \ref{sec:identification}.  For estimation and inference, we propose a simple two-step method of moments  procedure in Section \ref{sec:estimation}. We investigate the empirical performance of the proposed estimator through extensive simulation studies in Section \ref{sec:sim} and an application to evaluate the effect of self-efficacy on fatigue among health care workers during
the COVID-19 outbreak in Section \ref{sec:app}, before concluding with a brief discussion about possible extensions in Section \ref{sec:discussion}.

\section{Notation and assumptions}
 \label{sec:notation}

Suppose that $(O_1, \ldots, O_n)$ are independent and identically distributed observations of $O=(Y,A,\bm{M},X)$  from a population of interest, where $Y$ is the outcome, $A$ is the exposure, $\bm{M}=(M_1,...,M_K)^\T$ is a vector of $K$ mediators of interest for some $K\in \mathbb{Z}^+$ and $X$ is a vector of measured baseline covariates not affected by the exposure. \cite{robins1992identifiability} and \cite{pearl2001direct} defined what are now often called natural direct and indirect effects. To formalize these causal effects under potential outcomes framework \citep{neyman1923applications,rubin1974estimating}, let $Y({a,\mathbf{m}})$ denote the potential outcome that would be observed if $A=a$ and $\bm{M}=\mathbf{m}$. Similarly, let $\bm{M}(a)$ denote the potential mediator values that would be observed when exposure takes value $A=a$. Comparing any two values $a$, $a^{\prime}$ of the exposure, the population average natural direct and indirect effects through the $K$ mediators jointly are given by 
\begin{align}
NDE(a, a^{\prime})=E[Y(a,\bm{M}(a^{\prime}))-Y(a^{\prime},\bm{M}(a^{\prime}))],\quad NIE(a, a^{\prime})=E[Y(a,\bm{M}(a))-Y(a,\bm{M}(a^{\prime}))],
\end{align}
respectively \citep{vanderweele2014mediation}. The  average natural direct and indirect effects are particularly relevant for describing the
underlying mechanism by which the exposure operates, as their sum $NDE(a, a^{\prime})+NIE(a,a^{\prime})=E[Y(a,\bm{M}(a))-Y(a^{\prime},\bm{M}(a^{\prime}))]$ recovers the average total effect comparing any two values $a$, $a^{\prime}$ of the exposure. 

\cite{vanderweele2014mediation} proposed a parametric regression-based identification approach under the following assumptions that for any values $a$, $a^{\prime}$ of the exposure and $\mathbf{m}$ of the mediators,
\begin{itemize}
\item[A1$^\ast$.] $\bm{M}(a) \perp A \rvert X$; 
\item[A2$^\ast$.] $Y({a,\mathbf{m}}) \perp A \rvert X$;
\item[A3$^\ast$.] $Y({a,\mathbf{m}}) \perp \bm{M} \rvert A, X$;
\item[A4$^\ast$.] $Y({a,\mathbf{m}}) \perp \bm{M}(a^{\prime}) \rvert X$;
\item[A5$^\ast$.] $Y=Y({a,\mathbf{m}}) $ if $A=a$ and $\bm{M}=\mathbf{m}$, and $\bm{M}=\bm{M}(a)$ if $A=a$;
\item[A6$^\ast$.] $E(Y|A,\bm{M},X)=\bar{\theta}_0+\bar{\theta}_1 A +\bar{\theta}^\T_{2} \bm{M}+ \bar{\theta}^{\T}_{3} X$ and $E({M}_j|A,X)=\bar{\beta}_{0,j}+\bar{\beta}_{1,j}A+\bar{\beta}^{\T}_{2,j} X$ for $j=1,...,K$.
\end{itemize}
Assumptions (A1$^\ast$)--(A3$^\ast$) state that the vector of measured covariates $X$ is sufficiently rich to capture all confounding effects of the exposure-mediator, exposure-outcome and mediator-outcome relationships. Assumption (A4$^\ast$) is a cross-world independence assumption which, in conjunction with Assumption (A3$^\ast$), is equivalent to the absence of exposure-induced mediator-outcome confounders $L$ on a causal diagram interpreted as a nonparametric structural equation model with independent errors \citep{pearl2009causality}. If the variables in $L$ are measured then we can treat them as additional mediators in $\bm{M}$ and assess the mediation effects jointly \citep{vanderweele2014effect}. The potential outcomes are related to the observed data via the consistency Assumption (A5$^\ast$). Lastly, \cite{vanderweele2014mediation} considered the parametric outcome and mediator regressions in Assumption (A6$^\ast$) for simplicity and ease of estimation. Under Assumptions (A1$^\ast$)--(A6$^\ast$), \cite{vanderweele2014mediation} showed that the natural direct and indirect effects are encoded by the parameters
\begin{align}
\label{eq:param}
NDE(a, a^{\prime})=\bar{\theta}_1(a-a^{\prime}),\quad NIE(a, a^{\prime})=\bar{\beta}^\T_{1}\bar{\theta}_{2}(a-a^{\prime}),
\end{align}
respectively, where $\bar{\beta}_1=(\bar{\beta}_{1,1},...,\bar{\beta}_{1,K})^\T$.

\begin{figure}[!htbp]
	\vspace*{10pt}
	\centering	
	\begin{tikzpicture}[->,>=stealth',node distance=1cm, auto,]
		\node[est] (A) {$A$};
		\node[est, right=of A] (M1) {$M_1$};
		\node[est, right = of M1] (M2) { $M_2$};
		\node[est, right=of M2] (Y) {$Y$};
		\node[est, below=2cm of M2] (U) {$U$};
		\path[pil] (A) edgenode {} (M1);
		\path[pil] (A) edge [bend right=45] node {} (Y);
		\path[pil] (A) edge [bend right=45] node {} (M2);
		\path[pil] (M1) edgenode {} (M2);
		\path[pil] (M1) edge [bend left=45] node {} (Y);
		\path[pil] (M2) edgenode {} (Y);
		\path[pil] (U) edgenode {} (M1);
		\path[pil] (U) edgenode {} (M2);
		\path[pil] (U) edgenode {} (Y);
		
	\end{tikzpicture}
	\caption{Causal diagram with unmeasured mediator-outcome confounding  for two mediators within strata of $X$.}
	\label{fig1}
\end{figure}
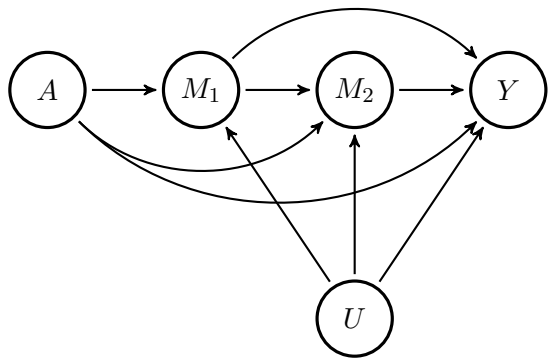

Although Assumptions (A1$^\ast$) and (A2$^\ast$) generally hold when the exposure is randomly assigned, in practical settings Assumption (A3$^\ast$) may be violated as it is often infeasible to also randomize or intervene on any of the $K$ mediators. To develop  the identification framework which allows for unmeasured mediator-outcome confounding, we follow \cite{ding2016sharp} and incorporate an additional set of unmeasured baseline covariates $U$ not affected by the exposure in the assumptions:
\begin{itemize}
\item[A1.] $\bm{M}(a) \perp A \rvert X,U$; 
\item[A2.] $Y({a,\mathbf{m}}) \perp A \rvert X,U$;
\item[A3.] $Y({a,\mathbf{m}}) \perp \bm{M} \rvert A,X,U$;
\item[A4.] $Y({a,\mathbf{m}}) \perp \bm{M}(a^{\prime}) \rvert X,U$;
\item[A5.] $Y=Y({a,\mathbf{m}}) $ if $A=a$ and $\bm{M}=\mathbf{m}$, and $\bm{M}=\bm{M}(a)$ if $A=a$;
\item[A6.] $E(Y|A,\bm{M},X,U)=\bar{\theta}_1 A +\bar{\theta}^\T_{2} \bm{M}+H(X,U)$ and $E({M}_j|A,X,U)=\bar{\beta}_{1,j}A+G_j(X,U)$ for $j=1,...,K$;
\item[A7.]  $U \perp A \rvert X$.
\end{itemize}
Assumptions  (A1)--(A4) are analogues of Assumptions (A1$^\ast$)--(A4$^\ast$) conditional on $(X,U)$. The partially linear regression models in Assumption (A6) extend the models in Assumption (A6$^\ast$) to incorporate $U$, where $H(X,U)$ and $G_j(X,U)$ are arbitrary functions of $(X,U)$ encoding their confounding effects on the outcome and $j$-th mediator, respectively. Assumption (A7) states that the exposure is essentially randomized within strata of $X$, either by design or through some natural experiments \citep{hernan2006instruments}. This scenario is depicted in Figure \ref{fig1} with two mediators. Based on the derivation in \cite{vanderweele2014mediation}, it is straightforward to show that under Assumptions (A1)--(A6) the natural direct and indirect effects are again given by (\ref{eq:param}). However, it is now more challenging to identify the parameters indexing the regressions in Assumption (A6) as they depend on unmeasured variables $U$. In the next section, we discuss how identification may be achieved in conjunction with Assumption (A7).

\section{Identification}

\label{sec:identification}

Under Assumptions (A6) and (A7), the observed data mediator regression models are 
\begin{equation}
\begin{split}
\label{eq:mediator}
E(M_j|A,X)=\bar{\beta}_{1,j} A+\bar{G}_{j}(X),\quad j=1,...,K, 
\end{split}
\end{equation}
where $\bar{G}_{j}(X)=E\{G_{j}(X,U)|A,X\}=E\{G_{j}(X,U)|X\}$ is an arbitrary function of the measured baseline covariates for $j=1,...,K$.  The $K$ separate mediator partially linear regression curves and hence the parameters $\bar{\beta}_{1}$ and $\bar{G}=(\bar{G}_{1},...,\bar{G}_{K})$ are identified from the observed data. It remains to identify the parameters $\bar{\theta}=(\bar{\theta}^\T_2,\bar{\theta}_1)^\T$ indexing the outcome regression.  \cite{fulcher2019estimation} proposed a higher moment condition to identify the natural indirect effect, which  can be extended under Assumptions (A6) and (A7) to yield the following  $K$ conditional mean independence conditions,
\begin{equation}
\begin{split}
\label{eq:ind1}
E\{(M_j-\bar{\beta}_{1,j}A -\bar{G}_{j}(X))(Y-\bar{\theta}^\T_{2} \bm{M})|A,X\}=\text{Cov}\{G_{j},H|A,X\}=\text{Cov}\{G_{j},H|X\},\quad j=1,...,K,
\end{split}
\end{equation}
where the dependencies of $G_j$, $H$ on $(X,U)$ for $j=1,...,K$ have been suppressed for brevity. While (\ref{eq:ind1}) suffices for identification of $\bar{\theta}_2$, it is generally  insufficient for identification of $\bar{\theta}$.  For example, when there are no measured baseline covariates and $A$ is binary, (\ref{eq:ind1}) places $K$  restrictions on the observed data law, but there are $K+1$ parameters in $\bar{\theta}$. As a remedy, we propose the following additional conditional mean independence condition  under Assumptions (A6) and (A7),
\begin{equation}
\begin{split}
\label{eq:ind2}
E(Y-\bar{\theta}_1 A -\bar{\theta}^\T_{2} \bm{M}|A,X)=E\{H(X,U)|A,X\}=E\{H(X,U)|X\}.
\end{split}
\end{equation}
The moment conditions in (\ref{eq:ind1}) and (\ref{eq:ind2}) can be summarized as
\begin{equation}
\begin{split}
\label{eq:ind3}
E\{\psi(O;\bar{\beta}_1,\bar{\theta},\bar{G})|A,X\}=E\{\psi(O;\bar{\beta}_1,\bar{\theta},\bar{G})|X\},
\end{split}
\end{equation}
where 
\begin{equation*}
	\psi(O;\bar{\beta}_1,\bar{\theta},\bar{G})=\begin{Bmatrix}
				(M_1-\bar{\beta}_{1,1} A-\bar{G}_{1}(X))(Y-\bar{\theta}^\T_{2} \bm{M})\\
		\vdots \\
		(M_K-\bar{\beta}_{1,K} A-\bar{G}_{K}(X))(Y-\bar{\theta}^\T_{2} \bm{M})\\
		Y-\bar{\theta}_1 A-\bar{\theta}^\T_{2} \bm{M}
	\end{Bmatrix}=\begin{Bmatrix}
		 (\bm{M}-E[ \bm{M}|A,X;\bar{\beta}_{1},\bar{G} ])(Y-\bar{\theta}^\T_{2} \bm{M})\\
		 	Y-\bar{\theta}_1 A-\bar{\theta}^\T_{2} \bm{M}
	\end{Bmatrix}.
\end{equation*}
 Following the G-estimation approach of \cite{robins1994correcting}, we consider the unconditional form of (\ref{eq:ind3}) for identification and estimation,
\begin{equation}
\begin{split}
\label{eq:g}
E[\{A-\bar{\pi}(X)\}\psi(O;\bar{\beta}_1,\bar{\theta},\bar{G})]=E[{\Psi}(O;\bar{\pi},\bar{\beta}_1,\bar{\theta},\bar{G})]=0,
\end{split}
\end{equation}
where $\bar{\pi}(X)=E(A|X)$ denotes the propensity score \citep{rosenbaum1983central} which is either known by design or identified from observed data.  To ensure that the solution to (\ref{eq:g}) is unique, we also require the following rank condition to hold \citep{rothenberg1971identification}:
\begin{itemize}
    \item[A8. ] The matrix $G={\partial E[{\Psi}(O;\bar{\pi},\bar{\beta}_1,{\theta},\bar{G})]}/{\partial \theta}$ is non-singular for all values of $\theta$.
\end{itemize}
By iterated expectations, $\text{det}(G)=(-1)^{K+1}G_{1} \text{det}(G_{2})$ where $G_{1}=E[\text{Var}(A|X)]$ and $G_{2}=E[\{A-\bar{\pi}(X)\}\text{Var}(\bm{M}|A,X)]$. This decomposition shows that $\text{det}(G)=0$ if either $\text{Var}(A|X)=0$ or $\text{Var}(\bm{M}|A,X)=\text{Var}(\bm{M}|X)$ almost surely. Therefore two necessary conditions for Assumption (A8) to hold are (1) $\text{Var}(A|X=x)>0$ for at least one value $x$ of the measured covariates and (2) $\text{Var}(\bm{M}|A=a,X=x^{\prime})\neq \text{Var}(\bm{M}|A=a^{\prime},X=x^{\prime})$ for at least one pair of exposure values $a\neq a^{\prime}$ and a value $x^{\prime}$ of the measured covariates. Condition (1) for identification of the natural direct effect parameter $\bar{\theta}_1$ is akin to the positivity assumption for a binary exposure in causal inference under unconfoundedness conditions and typically holds when the exposure is randomly assigned by the investigator. Condition (2) leverages heteroscedasticity of the mediators with respect to the exposure for identification of $\bar{\theta}_2$, which is plausible under a variety of settings in the health and social sciences due to heterogeneity in the exposure's effect on the mediators.  For example, consider the following mediator structural equation models \citep{pearl2009causality}
$$M_j =\beta_{1,j}(\varepsilon_j)A+G_{j}(X,U)  ,\quad j=1,...,K,$$
where $\varepsilon_j\perp A,X,U$ represents heterogeneity in the effect of $A$ on $M_j$.  \cite{tchetgen2021genius} showed that $\text{Var}(M_j|A,X)$ remains a function of $A$ provided that the distribution of $\varepsilon_j$ is non-degenerate, for $j=1,...,K$. 

\begin{lemma}
\label{lem:1}
Under Assumptions (A6)--(A8), the parameters $\bar{\beta}_1$ and $\bar{\theta}$ are identified from the observed data.
\end{lemma}

\section{A two-step method of moments estimation approach}

\label{sec:estimation}

The semiparametric identification result in Lemma \ref{lem:1} involves the nonparametric nuisance components $\bar{\pi}$ and $\bar{G}$.  In principle it is possible to estimate these components nonparametrically under sufficient smoothness conditions. However, if $X$ contains numerous continuous variables, the resulting estimators may exhibit poor finite sample behavior  in moderately sized samples as the data are too sparse to conduct stratified estimation \citep{robins1997toward}. For this reason as well as simplicity and ease of implementation, in this paper we propose a two-step method of moments approach in which $\bar{\pi}$ and $\bar{G}$ are estimated under the following parametric assumptions in the first step,
\begin{itemize}
    \item[A9.] $\bar{\pi}(X)=\pi(C;\bar{\gamma})$ and $\bar{G}_j(X)={G}_j(X;\bar{\alpha}_j)$ for  $j=1,...,K$, where $\pi(C;\gamma)$ and ${G}_j(X;{\alpha}_j)$ are known functions smooth in the finite-dimensional parameters $\gamma$ and ${\alpha}_j$ respectively.  
\end{itemize}
Throughout let  $  \widehat{E}_n\{g(O)\}=n^{-1}\sum_{i=1}^n g(O_i)$ denote the empirical mean operator and ${\alpha}=(\alpha^\T_1,...,\alpha^\T_K)^\T$.

\subsection{ First step}
Let $S(A,C;\gamma)$ and $S(M,A,X;\alpha,\beta_1)$ denote some user-specified unbiased estimating functions for the parameters $\bar{\gamma}$, $\bar{\alpha}$ and $\bar{\beta}_1$ respectively based on Assumption (A9). For example, typically a logistic regression model is used for binary $A$, say $\pi(X;\gamma)=\{1+\exp(-\gamma^\T X)\}^{-1}$, in which case the score function of $\gamma$ is 
\begin{equation}
\label{eq:s1}
S(A,X;\gamma)=X\{A-\pi(X;\gamma)\}.
\end{equation}
In addition, if we specify the linear model $G_j(X;\alpha_j)=\alpha_{j}^\T X$ for the $j$-th mediator regression, $j=1,...,K$, then 
\begin{equation}
\label{eq:s2}
S(M,A,X;\alpha,\beta_1)=\begin{Bmatrix}
				(A,X^\T)^\T(M_1-\beta_{1,1} A-\alpha_{1}^\T X)\\
		\vdots \\
		(A,X^\T)^\T(M_K-\beta_{1,K} A-\alpha_{K}^\T X)
	\end{Bmatrix}.
\end{equation}	
Let $\widehat{\gamma}$, $\widehat{\alpha}$ and $\widehat{\beta}_1$ denote the estimators which jointly solve the empirical moment condition
$$ \widehat{E}_n\{S^{\T}(A,C;\gamma),S^\T(M,A,X;\alpha,\beta_1)\}^{\T}=0,$$
and let $\widehat{\pi}(X)=\pi(X;\widehat{\gamma})$ and $\widehat{G}(X)=\{G_1(X;\widehat{\alpha}_1),...,G_K(X;\widehat{\alpha}_K)\}$. 
 
\subsection{ Second step}

 In the second stage, we obtain $\widehat{{\theta}}$ which solves the following empirical version of moment condition (\ref{eq:g}), $$ \widehat{E}_n[{\Psi}(O;\widehat{\pi},\widehat{\beta}_1,{\theta},\widehat{G})]=0.$$
 An appealing feature of the two-step procedure is that the nuisance components $\bar{\pi}$ and $\bar{G}$ can be estimated before the second stage involving the outcome data, and therefore mitigates potential for "data-dredging" exercises that comes with a fully specified outcome model \citep{rubin2007design}.

\subsection{ Inference}
For inference, the empirical moments  involved in the two-step procedure can be ``stacked'' to account for the variability associated with the first-step estimation of nuisance parameters \citep{newey1994large}. Specifically, let $\varphi=({\gamma}^\T,{\alpha}^\T,{\beta}^\T_1,\theta^\T)^\T$. Then the estimator $\widehat{{\varphi}}$ may be viewed as solving the joint empirical moment condition $\widehat{E}_n\{\Phi(O;{\varphi})\}=0$, where $$\Phi(O;{\varphi})= \{S^{\T}(A,C;\gamma),S^\T(M,A,X;\alpha,\beta_1),{\Psi}^\T(O;\pi(\gamma),{\beta}_1,{\theta},G(\alpha))\}^\T.$$ The following result on asymptotic normality of $\widehat{{\varphi}}$ follows from standard M-estimation theory \citep{newey1994large}.
 
\begin{lemma}
Suppose Assumptions (A6)--(A9) hold. Then under standard regularity conditions and as $n\to\infty$, $$\sqrt{n}(\widehat{{\varphi}}-\bar{{\varphi}}) \xrightarrow{D}  N\left(0, \left[\frac{\partial E\{\Phi(O;{\varphi})\}}{\partial \varphi}\biggr \rvert_{\varphi=\bar{\varphi}}\right]^{-1} E\{\Phi(O;\bar{\varphi})\Phi^\T(O;\bar{\varphi})\}\left[\frac{\partial E\{\Phi(O;{\varphi})\}}{\partial \varphi}\biggr \rvert_{\varphi=\bar{\varphi}}\right]^{-1\T}\right).$$ 
\end{lemma}
\noindent It follows that, in conjunction with  Assumptions (A1)--(A5), the estimators of the natural direct and indirect effects per unit change in the exposure are given by
\begin{align}
\label{eq:param2}
\widehat{NDE}=\widehat{\theta}_1,\quad \widehat{NIE}=\widehat{\beta}^\T_{1}\widehat{\theta}_{2},
\end{align}
respectively, and their limiting variances can be derived using the multivariate delta method. Alternatively, nonparametric bootstrap may
also be used to perform inference.

\section{Simulation study}
\label{sec:sim}

In order to investigate the numerical performance of the proposed estimators, we perform Monte Carlo simulations involving independent and identically distributed data $(Y,A,\bm{M},X,U)$ with three mediators $\bm{M}=(M_1,M_2,M_3)^\T$ generated as follows:
\begin{align*}
&X\sim N(0,1);\quad  U\mid X \sim N(1+0.5 X,1),\\
&A\mid X,U \sim {Bernoulli}\left\{p=\frac{1}{1+\exp(-0.8-1.2 X)}\right\},\\
&\bm{M}\mid A,X,U \sim  N
\begin{Bmatrix}
\begin{pmatrix}
1.2+1.5 A+1.1 X+\eta U \\
0.5+1.2 A+1.8 X+\eta U \\
1.3+1.0 A+0.5 X+\eta U 
\end{pmatrix} ,
\bm{\Sigma}
\end{Bmatrix},\\
& Y \mid A,\bm{M},X,U \sim N(1.3+2.5 A +1.2 M_1+0.8 M_2 +M_3+1.5 X +\eta U,1),
\end{align*} 
where $\bm{\Sigma}$ has the Toeplitz covariance structure $\text{Cov}(M_i,M_j)=2^{-|i-j|}(1+\delta A)$ for $1\leq i,j\leq 3$. Under this data generating mechanism, we vary (i) the sample size $n=400$ or $800$, (ii) the degree of unmeasured mediator-outcome confounding encoded by $\eta=0$ or $0.5$ as well as (iii) the magnitude of heteroscedasticity encoded by $\delta=2.0$ or $5.0$. We compare the proposed estimators $\widehat{NDE}$ and $\widehat{NIE}$ implemented using the correctly specified models (\ref{eq:s1}) and (\ref{eq:s2}) with the regression-based estimators $\widehat{NDE}_{reg}$ and $\widehat{NIE}_{reg}$ of \cite{vanderweele2014mediation}. Standard errors are obtained using the empirical sandwich estimator \citep{newey1994large}. 

The following remarks can be made based on the results of 1000 simulation replicates summarized in Tables \ref{tab:1} and \ref{tab:2}. The proposed estimators $\widehat{NDE}$ and $\widehat{NIE}$ show negligible bias and coverage proportion close to the nominal level of $0.95$ once sample size reaches $n=800$ across all simulation settings, but with substantially higher variance than $\widehat{NDE}_{reg}$ and $\widehat{NIE}_{reg}$. In agreement with theory,  $\widehat{NDE}_{reg}$ and $\widehat{NIE}_{reg}$ perform well in the absence of unmeasured outcome-mediator confounding but exhibit noticeable bias and undercoverage otherwise. The variance of the proposed estimators generally decreases when sample size or the magnitude of mediator heteroscedasticity increases.

  \begin{table}

		\begin{center}
		\caption{Comparison of methods for estimation of the natural direct and indirect effects in the absence of unmeasured outcome-mediator confounding $(\eta=0)$. The
two rows of results for each estimator correspond to sample sizes of $n = 400$ and $n = 800$
respectively.}
	\label{tab:1}
		\bigskip
	
		\begin{tabular}{cccccccccccc}
			\toprule
	    & $\widehat{NDE}$ & $\widehat{NIE}$ & $\widehat{NDE}_{reg}$ &$\widehat{NIE}_{reg}$\\
	     \hline\noalign{\smallskip}
&  \multicolumn{4}{c}{$\delta=2.0$}\\
$|\text{Bias}|$ & 0.014 & 0.021& 0.007 & 0.000  \\
                      & 0.008 & 0.008 & 0.001 & 0.000  \\
$\sqrt{\text{Var}}$ & 0.268 & 0.415 & 0.134 & 0.378   \\   
                             &0.178 & 0.267 & 0.095 & 0.253    \\                
$\sqrt{\text{EVar}}$ & 0.422 & 0.422 & 0.375 & 0.375 \\
                               & 0.177 & 0.287 & 0.091 & 0.265  \\
Cov95 & 0.959 & 0.947 & 0.937 & 0.945 \\
          & 0.956 & 0.962 & 0.941 & 0.956\\

&  \multicolumn{4}{c}{$\delta=5.0$}\\
$|\text{Bias}|$ & 0.005 & 0.010& 0.007 & 0.001  \\
                      & 0.005 & 0.005 & 0.001 & 0.000  \\
$\sqrt{\text{Var}}$ & 0.167 & 0.471 & 0.129 & 0.483   \\   
                             &0.116 & 0.316 & 0.091 & 0.328    \\                
$\sqrt{\text{EVar}}$ & 0.164 & 0.471 & 0.124 & 0.482 \\
                               & 0.113 & 0.332 & 0.088 & 0.340  \\
Cov95 & 0.948 & 0.947 & 0.938 & 0.945 \\
          & 0.950 & 0.962 & 0.941 & 0.954\\
\hline
		\end{tabular}
    \begin{tablenotes}
      \item{\noindent \small Note: $|\text{Bias}|$ and $\sqrt{\text{Var}}$ are the Monte Carlo absolute bias and standard deviation of the point estimates,  $\sqrt{\text{EVar}}$ is the
square root of the mean of the variance estimates and Cov95 is the coverage proportion of
the 95\% confidence intervals, based on 1000 repeated simulations. Zeros denote values smaller than $0.0005$.}
   \end{tablenotes}
	\end{center}
  \end{table}

 \begin{table}

		\begin{center}
		\caption{Comparison of methods for estimation of the natural direct and indirect effects in the presence of unmeasured outcome-mediator confounding $(\eta=0.5)$. The
two rows of results for each estimator correspond to sample sizes of $n = 400$ and $n = 800$
respectively.}
	\label{tab:2}
		\bigskip
	
		\begin{tabular}{cccccccccccc}
			\toprule
	    & $\widehat{NDE}$ & $\widehat{NIE}$ & $\widehat{NDE}_{reg}$ &$\widehat{NIE}_{reg}$\\
	     \hline\noalign{\smallskip}
&  \multicolumn{4}{c}{$\delta=2.0$}\\
$|\text{Bias}|$ & 0.053 & 1.195& 0.198 & 1.446  \\
                      & 0.016 & 0.034 & 0.194 & 0.176  \\
$\sqrt{\text{Var}}$ & 0.384 & 0.511 & 0.146 & 0.434   \\   
                             &0.230 & 0.318 & 0.101 & 0.296    \\                
$\sqrt{\text{EVar}}$ & 0.454 & 0.574 & 0.140 & 0.434 \\
                               & 0.231 & 0.340 & 0.099 & 0.307  \\
Cov95 & 0.962 & 0.953 & 0.697 & 0.926 \\
          & 0.955 & 0.961 & 0.496 & 0.923\\

&  \multicolumn{4}{c}{$\delta=5.0$}\\
$|\text{Bias}|$ & 0.012 & 1.238& 0.119 & 1.368  \\
                      & 0.003 & 0.021 & 0.115 & 0.097  \\
$\sqrt{\text{Var}}$ & 0.194 & 0.501 & 0.142 & 0.526  \\   
                             &0.136 & 0.339 & 0.099 & 0.361    \\                
$\sqrt{\text{EVar}}$ & 0.194 & 0.506 & 0.136 & 0.528 \\
                               & 0.133 & 0.356 & 0.096 & 0.373  \\
Cov95 & 0.948 & 0.952 & 0.839 & 0.943 \\
          & 0.951 & 0.952 & 0.763 & 0.947\\
\hline
		\end{tabular}
    \begin{tablenotes}
      \item{\noindent \small Note: See the footnote of Table \ref{tab:1}.}
   \end{tablenotes}
	\end{center}
  \end{table}
\section{Application}
\label{sec:app}

We apply the proposed methods in reanalysing an observational dataset from \cite{Houdata} on $n=527$ healthcare workers from Anhui Province, China during the COVID-19 outbreak in March 2020 to evaluate the mediating effects of post-traumatic stress disorder symptoms in the association between self-efficacy and fatigue.  We refer interested readers to \cite{hou2020self} for further details on the study design. The binary exposure $A$ is the total score on the General Self-Efficacy Scale dichotomized at the sample median while the outcome of interest $Y$ is the standardized total score on the 14-item Fatigue Scale. The three mediators under consideration ($M_1,M_2,M_3$) are standardized scores in post-traumatic stress disorder symptoms reported in three categories; re-experiencing, avoidance and hyperarousal. The vector of measured baseline variables $X$ include age, level of negative coping, gender, marital status, education level, years of working experience and technical title.

We implement the proposed estimators $\widehat{NDE}$ and $\widehat{NIE}$ using generalized linear models (\ref{eq:s1}) and (\ref{eq:s2}) together with the regression-based estimators $\widehat{NDE}_{reg}$ and $\widehat{NIE}_{reg}$ of \cite{vanderweele2014mediation} to estimate the natural direct and indirect effects of self-efficacy on fatigue mediated though the three categories of  post-traumatic stress disorder symptoms jointly. Table \ref{tab3} shows the analysis results. The regression-based estimators of \cite{vanderweele2014mediation} have noticeably smaller standard errors than the proposed estimators, in agreement with the Monte Carlo results, and yield statistically significant  natural direct and indirect effects at $\alpha$-level of 0.05. On the other hand, the point estimate of $\widehat{NIE}$ is close to zero and statistically insignificant at the same $\alpha$-level. This result suggests that we cannot rule out a null joint indirect effect after accounting for potential unmeasured confounding of the mediator-outcome relationship.

 \begin{table}
\centering

\caption{Estimates of direct and indirect effects of self-efficacy on fatigue mediated though three post-traumatic stress disorder symptoms jointly.}{%
\begin{tabular}{ccccccccccccc}
 \toprule
  $\widehat{NDE}$ & $\widehat{NIE}$ & $\widehat{NDE}_{reg}$ &$\widehat{NIE}_{reg}$ \\
	     \midrule
  $-0.735\pm 0.486$  & $0.040\pm 0.474$& $-0.405\pm 0.145$ & $-0.289\pm 0.088$\\
\bottomrule
\end{tabular}}
\label{tab3}

   \begin{tablenotes}
      \small
      \item Note: Estimate $\pm$ $1.96\times$standard error. 
    \end{tablenotes}
    
\end{table}

\section{Discussions}
\label{sec:discussion}

One of the main concerns with causal mediation analysis with multiple mediators is the inability to categorically rule out the existence of unmeasured mediator-outcome confounders, as it is generally infeasible to randomize any of the mediators under consideration. In this paper, we build upon the works of \cite{vanderweele2014mediation} and \cite{fulcher2019estimation} to develop a simple two-step method of moments estimation procedure to assess the joint mediation effects of multiple mediators  in the presence of potential unmeasured mediator-outcome confounding. Our work can be extended in several important directions, including identification and estimation of path-specific effects \citep{10.5555/1642293.1642350,albert2011generalized,daniel2015causal} or their interventional analogues  \citep{vansteelandt2017interventional,lin2017interventional,loh2020heterogeneous} as well as causal mediation analysis of survival outcome with multiple mediators \citep{huang2017causal,lin2017mediation}, which we plan to pursue in future research.

\section*{Acknowledgement}
Baoluo Sun is supported by the National University of Singapore Start-Up Grant (R-155-000-203-133).

	\thispagestyle{empty}
	\bibliographystyle{apalike}
	\bibliography{refs}

	\clearpage

\end{document}